\documentclass[a4paper,aps,prd,nofootinbib,onecolumn,notitlepage]{revtex4-1}
\RequirePackage[english]{babel}
\RequirePackage[latin1]{inputenc}
\RequirePackage[T1]{fontenc}
\RequirePackage{mathrsfs}
\RequirePackage{amsmath}
\RequirePackage{amssymb}
\RequirePackage{amsbsy}
\RequirePackage{bm}
\usepackage[lofdepth,lotdepth]{subfig}
\usepackage{graphicx}
\usepackage{multirow} 
\usepackage{caption}
\def\de#1/de#2{\frac{\partial {#1}}{\partial {#2}}}

\newcommand{\nn}{\nonumber}
\newcommand{\ba}{\begin{eqnarray}}
\newcommand{\ea}{\end{eqnarray}}
\newcommand{\be}{\begin{equation}}
\newcommand{\ee}{\end{equation}}

\begin{document}
\title{Relativistic polytropic equations of state in Ho\v rava gravity and Einstein-\ae ther theory}
\author{Daniele Vernieri}
\affiliation{Centro de Astrof\'isica e Gravita\c c\~ao - CENTRA,
Departamento de F\'{\i}sica, Instituto Superior T\'ecnico - IST,
Universidade de Lisboa - UL, Avenida Rovisco Pais 1, 1049-001,
Portugal}
\date{\today}
\begin{abstract}
The equations of state for a characteristic spacetime are studied in the context of the spherically symmetric interior exact and analytical solutions in Ho\v rava gravity and Einstein-\ae ther theory in which anisotropic fluids are considered. In particular, for a given anisotropic interior solution, the equations of state relating the density to the radial and tangential pressure are derived, by means of a polynomial best-fit. Moreover, the well-known relativistic polytropic equations of state are used in order to obtain the profile of the thermodynamical quantities inside the stellar object as provided by the specific exact solution considered. It is then shown that these equations of state need to be modified in order to account for the profiles of density and pressures. 
\end{abstract}
\maketitle

\section{Introduction} 

Ho\v rava gravity was proposed in 2009 as a candidate theory for quantum gravity~\cite{Horava:2009uw,Blas:2009qj}. Since then a lot of work has been done in the attempt to prove that the theory is renormalizable beyond the power-counting arguments~\cite{Barvinsky:2015kil,Bellorin:2016wsl,Barvinsky:2017zlx}. We will not consider here any of the restricted versions of Ho\v rava gravity that have been proposed in order to reduce the number of independent couplings~\cite{Sotiriou:2009gy,Weinfurtner:2010hz,Vernieri:2011aa,Vernieri:2012ms,Vernieri:2015uma}.

The theory is by construction built in a preferred foliation. This means that a preferred direction is defined in any point of the spacetime and it is intended to be the one toward which the preferred time flows. Then, Lorentz invariance is not an exact symmetry and it is broken at any energy scale. Moreover, an extra scalar graviton propagates besides the spin-2 mode. This fact is quite expected since in general less symmetry implies more degrees of freedom.

The structure of the field equations of Ho\v rava gravity is generically highly non-linear, even at low-energies that we will consider in the following. This is a stumbling block, specially if one is interested in studying the phenomenological implications of the theory without relying on numerics. 
Indeed, the exact analytical solutions that have been found up to now in the IR limit of the most general version of Ho\v rava gravity, {\it e.g.} the ones related to black holes, are very few, and they are obtained by considering very specific tuned choices of the parameters in the gravitational action~\cite{Berglund:2012bu} or working in three-dimensional spacetimes~\cite{Sotiriou:2014gna}. Moreover, other exact black hole solutions have been obtained in some restricted versions of the theory with the inclusion of higher-order operators~\cite{Lu:2009em,Cai:2009pe,Park:2009zra}. On the contrary, the only solutions describing the interior of neutron stars by means of perfect isotropic fluids are all numerical~\cite{Eling:2006df,Eling:2007xh}. 
The generic difficulty in finding exact solutions to the field equations, regardless of the background, is one of the main reasons for which the reconstruction algorithm implemented in Ref.~\cite{Vernieri:2017dvi} is so relevant. Indeed, by means of the approach described therein, it is possible to find viable exact and analytical interior solutions to the field equations in spherically symmetric spacetimes with the presence of anisotropic fluids~\cite{Herrera:1997plx}. In particular, it is shown that the system of independent equations can be exactly solved by choosing the interior structure of the spacetime, {\it i.e.}, by properly selecting the metric coefficients. By using the field equations, it is then possible to find algebraically the explicit analytic expressions for the thermodynamical quantities, which are the density and the radial and transversal pressure. With the exact solutions at hand, one can {\it a posteriori} verify that they satisfy all the relevant physical requirements~\cite{Harko:2002db} that make them viable in order to properly describe the interior of astrophysical objects.

One of the most important features of this method, is that the equation of state (EoS) of the inner fluid is left unspecified, and this seems to be very realistic. Indeed, although many theoretical approaches have been developed to find a proper effective description of the stars interior~\cite{Ozel:2016oaf}, the characterizing EoS is still unknown. Moreover, we also expect that in a quantum theory like Ho\v rava gravity the physical properties of the fluid might be much different with respect to the ones expected in a standard scenario like, {\it e.g.}, in general relativity (GR) because of the intrinsically modified description of the gravitational interaction.
Furthermore, by means of the reconstruction algorithm and the resulting exact solutions that have been derived, it will be possible to get the analytic form of the EoS {\it a posteriori}, as we will discuss in detail. In this Paper we will focus on the specific set of viable solutions found in Ref.~\cite{Vernieri:2017dvi}, and we will study the EoS that can be deduced from there. 

In Sec.~\ref{sec1} we introduce the action of Ho\v rava gravity and Einstein-\ae ther theory~\cite{Jacobson:2000xp} and discuss their equivalence when the \ae ther is taken to be hypersurface-orthogonal at the level of the action.
In Sec.~\ref{sec2} the field equations are derived by considering a spherically symmetric background with the addition of anisotropic fluids.
In Sec.~\ref{sec3} the main features of the exact solutions considered are discussed. In Sec.~\ref{sec4} the EoS for the resulting fluid quantities is studied. We first perform a polynomial fit in order to find the EoS relating the density to the radial and tangential pressure. Then, we use different relativistic polytropic EoS and show that the relativistic models widely used in the literature~\cite{Nilsson:2000zg,Herrera:2013fja,Herrera:2014caa,Isayev:2017rci} need to be modified. In Sec.~\ref{sec5} the conclusions are discussed.    
 
\section{Ho\v rava Gravity}
\label{sec1}

Let us start by writing down the most general action of Ho\v rava gravity which, in the preferred foliation, looks like
\be \label{horava}
\mathcal{S}_{H}=\frac{1}{16\pi G_H}\int{dT d^3x\sqrt{-g}\left(K_{ij}K^{ij}-\lambda K^2 +\xi \mathcal{R}+\eta a_i a^i+\frac{\mathcal{L}_4}{M_\ast^2}+\frac{\mathcal{L}_6}{M_\ast^4}\right)}+S_m[g_{\mu\nu},\psi],
\ee 
where $G_H$ is the effective gravitational coupling constant, $g$ is the determinant of the metric $g_{\mu\nu}$, $\mathcal{R}$ is the Ricci scalar of the three-dimensional constant-$T$ hypersurfaces, $K_{ij}$ is the extrinsic curvature, $K$ is its trace, and $a_i=\partial_i \mbox{ln} N$, where $N$ is the lapse function. The couplings $\left\{\lambda,\xi,\eta\right\}$ are dimensionless, and in order to recover GR, they must take identically the values $\left\{1,1,0\right\}$. Moreover, $\mathcal{L}_4$ and $\mathcal{L}_6$ collectively denote the fourth-order and sixth-order operators, respectively, and $M_\ast$ is the mass scale suppressing them. Finally, $S_m$ is the matter action for the matter fields collectively denoted by $\psi$. 
In the following, we only consider the low-energy limit of the theory, which amounts to discarding the higher-order operators contained in $\mathcal{L}_4$ and $\mathcal{L}_6$. Also, in the forthcoming sections, we will only consider the covariantized version of the low-energy limit of Ho\v rava gravity, which is referred to as the {\it khronometric} model. To make use of that, let us first take into account the action of Einstein-\ae ther theory~\cite{Jacobson:2000xp}, which is
\be
\mathcal{S}_{\mbox{\footnotesize \ae}}=-\frac{1}{16\pi G_{\mbox{\footnotesize \ae}}}\int{d^4x\sqrt{-g}\left(R+\mathcal{L}_{\mbox{\scriptsize\ae}}\right)}+S_m[g_{\mu\nu},\psi], \label{aether}
\ee
where $G_{\mbox{\footnotesize \ae}}$ is the ``bare'' gravitational constant and $\mathcal{L}_{\mbox{\scriptsize\ae}}$ is given by
\be
\mathcal{L}_{\mbox{\footnotesize \ae}}=c_1\nabla^\alpha u^\beta\nabla_\alpha u_\beta+c_2\nabla_\alpha u^\alpha\nabla_\beta u^\beta+c_3\nabla_\alpha u^\beta\nabla_\beta u^\alpha+c_4 u^\alpha u^\beta\nabla_\alpha u_\nu\nabla_\beta u^\nu\,,
\ee
where the $c_i$ are arbitrary constant coefficients and $u^\mu$ is a unit timelike vector field, {\it i.e.} $g_{\mu\nu}u^\mu u^\nu=1$, which is called the ``\ae ther'' vector field. To show the relation between Ho\v rava gravity and Einstein-\ae ther theory let us consider the \ae ther to be hypersurface-orthogonal at the level of the action, which locally amounts to choosing
\be
u_\alpha=\frac{\partial_\alpha T}{\sqrt{g^{\mu\nu}\partial_\mu T \partial_\nu T}}\,,
\ee
where the preferred time $T$ becomes a scalar field (referred to as the ``{\it khronon}'') in the covariant formulation, which defines the preferred foliation. The two actions in Eqs.~\eqref{horava} and~\eqref{aether} can then be mapped into each other if the following relations among the parameters hold~\cite{Jacobson:2010mx}:
\be
	\label{eqn:corresp}
\frac{G_H}{G_{\scriptsize\mbox{\ae}}}=\xi = \frac{1}{1-c_{13}}\,, \hspace{2em} \frac{\lambda}{\xi} = 1 + c_2\,, \hspace{2em} \frac{\eta}{\xi} = c_{14}\,,
\ee
where $c_{ij} = c_i+c_j$. In the following, we undertake for convenience the covariant approach. 

\section{Spherical Symmetry and Anisotropic Fluids}
\label{sec2}

Let us now define all the relevant quantities that will be useful in what follows. 
By assuming spherical symmetry, the metric can be written in full generality as
\be
ds^2 = A(r) dt^2 - B(r) dr^2 - r^2\,\big(d\theta^2+\sin^2\theta d\phi^2\big). \label{Eq0}
\ee
Moreover, we consider an anisotropic fluid whose stress-energy tensor is given by 
\be
T_{\mu\nu}=\left(\rho + p_t\right) v_\mu v_\nu - p_t g_{\mu\nu} + \left(p_r-p_t\right) s_\mu s_\nu,
\ee
where $\rho$ is the density, $p_r$ and $p_t$ are the radial and transversal pressure, respectively, $v^\mu$ denotes the 4-velocity of the fluid
\be
v^\mu =\biggl(\frac{1}{\sqrt{A(r)}},0,0,0\biggr),
\ee
and $s^\mu$ is a spacelike 4-vector defined as
\be
s^\mu = \biggl(0,\frac{1}{\sqrt{B(r)}},0,0\biggr),
\ee
with the properties $s^\mu s_\mu = -1$ and $s^\mu u_\mu = 0$.
It is easy to show that the components of the stress-energy tensor are given by
\be
T_\mu^{\phantom{\mu}\nu} = \mbox{diag}\big(\rho,-p_r,-p_t,-p_t\big).
\ee 
The \ae ther vector field, which is timelike and by definition normalized to $1$, in spherical symmetry is always hypersurface-orthogonal and takes the general form: 
\be
u^\alpha =\biggl(F(r),\sqrt{\frac{A(r) F(r)^2-1}{B(r)}},0,0\biggr), 
\ee
where $F(r)$ is a generic function. 
Nevertheless, here, we take into account for simplicity a static \ae ther $u^\mu$~\cite{Eling:2006df,Eling:2007xh,Vernieri:2017dvi} given by
\be
u^\alpha =\biggl(\frac{1}{\sqrt{A(r)}},0,0,0\biggr),
\ee
which results in being aligned with the fluid velocity $v^\mu$. 

The independent field equations that we have to consider are:
\be
\frac{\eta }{ \xi }\left[-\frac{A''(r)}{2  A(r) B(r)}+\frac{ A'(r) B'(r)}{4  A(r) B(r)^2}+\frac{3   A'(r)^2}{8  A(r)^2 B(r)}-\frac{ A'(r)}{r A(r) B(r)}\right]+\frac{B'(r)}{r B(r)^2}-\frac{1}{r^2 B(r)}+\frac{1}{r^2}=8 \pi G_{\mbox{\footnotesize \ae}} \rho (r)\,, \label{Eq1}
\ee 
\be
\frac{\eta A'(r)^2}{8 \xi  A(r)^2 B(r)}+\frac{A'(r)}{r A(r) B(r)}+\frac{1}{r^2 B(r)}-\frac{1}{r^2}=8 \pi G_{\mbox{\footnotesize \ae}} p_r(r)\,, \label{Eq2}
\ee
\be
-\frac{\eta  A'(r)^2}{8 \xi A(r)^2 B(r)}+\frac{A''(r)}{2 A(r) B(r)}-\frac{A'(r) B'(r)}{4 A(r) B(r)^2}+\frac{A'(r)}{2 r A(r) B(r)}-\frac{A'(r)^2}{4 A(r)^2 B(r)}-\frac{B'(r)}{2 r B(r)^2}=8 \pi G_{\mbox{\footnotesize \ae}} p_t(r)\,, \label{Eq3}
\ee
which are, respectively, the modified Einstein equations $(0-0)$, $(1-1)$, and $(2-2)$. Furthermore, one has to consider also the conservation equation for the stress-energy tensor, {\it i.e.} $\nabla^\mu T_{\mu\nu}=0$, which can be written as:
\be
p_r'(r)+\left[\rho(r)+p_r(r)\right]\frac{A'(r)}{2A(r)}=\frac{2}{r} \left[p_t(r)-p_r(r)\right]. \label{conserv}
\ee
However, only three of the equations above are independent. Indeed, it can be easily shown that writing $\rho(r)$, $p_r(r)$ and $p_t(r)$ in terms of the metric coefficients by means of Eqs.~\eqref{Eq1},~\eqref{Eq2} and~\eqref{Eq3}, respectively, then the conservation Eq.~\eqref{conserv} is automatically satisfied once the formers have been substituted. Notice that, because of spherical symmetry, the field equations above identically coincide with the ones obtained in the framework of Einstein-\ae ther theory~\cite{Blas:2010hb}. So, the spherically symmetric solutions to the low-energy limit of Ho\v rava gravity and Einstein-\ae ther theory identically coincide. 

\section{A Viable Interior Solution}
\label{sec3}

Let us now consider the viable solution found in Ref.~\cite{Vernieri:2017dvi} by means of the reconstruction algorithm described therein. Indeed, due to the structure of the field equations,  it is possible to generate a double infinity of exact analytical solutions. The procedure basically consists in fixing suitably the metric coefficients $A(r)$ and $B(r)$, and then solve the equations in order to find $\rho$, $p_r$ and $p_t$.
We just give here the expressions for the metric coefficients, which are qualitatively similar to the metric coefficients of the Tolmann IV solution for an isotropic fluid in GR~\cite{Tolman:1939jz}, and the thermodynamical quantities which solve the field equations and lead to a viable physical solution as already shown in detail in Ref.~\cite{Vernieri:2017dvi}:
\be
A(r)=D_1+D_2 r^2\,, \,\,\,\,\,\, B(r)=\frac{D_3+D_4 r^2}{D_3+D_5 r^2+D_6 r^4}\,, \label{Eq4}
\ee
\ba
\rho(r) &=& -\frac{1}{16 \pi G_{\mbox{\footnotesize \ae}} \xi \left(D_1+D_2 r^2\right)^2 \left(D_3+D_4 r^2\right)^2}\left\{-2 D_1^2 \xi  \left[D_3 \left(3 D_4-3 D_5-5 D_6 r^2\right)+D_4 r^2 \left(D_4-D_5-3 D_6 r^2\right)\right]\right. \nn \\
&&+2 D_1 D_2 \left[3 D_3^2 \eta +D_3 r^2 \left(2 D_4 (\eta -3 \xi )+D_5 (4 \eta +6 \xi )+5 D_6 r^2 (\eta +2 \xi )\right)+D_4 r^4 \left(-2 D_4 \xi +D_5 (3 \eta +2 \xi ) \right.\right. \label{Eq7} \nn \\
&&\left.\left.+2 D_6 r^2 (2 \eta +3 \xi )\right)\right]+D_2^2 r^2 \left[3 D_3^2 \eta +D_3 r^2 \left(D_4 (\eta -6 \xi )+D_5 (5 \eta +6 \xi )+D_6 r^2 (7 \eta +10 \xi )\right) \right. \nn \\
&&\left.\left. +D_4 r^4 \left(-2 D_4 \xi +D_5 (3 \eta +2 \xi )+D_6 r^2 (5 \eta +6 \xi )\right)\right]\right\}, \label{Eq8} \nn \\
\ea
\ba
p_r(r)&=&\frac{1}{16 \pi G_{\mbox{\footnotesize \ae}} \xi \left(D_1+D_2 r^2\right)^2 \left(D_3+D_4 r^2\right)} \left\{2 D_1^2 \xi  \left(-D_4+D_5+D_6 r^2\right)+4 D_1 D_2 \xi \left[D_3+r^2 \left(2 \left(D_5+D_6 r^2\right)-D_4\right)\right]\right. \nn \\
&&\left.+D_2^2 r^2 \left[D_3 (\eta +4 \xi )+r^2 \left(-2 D_4 \xi +D_5 (\eta +6 \xi )+D_6 r^2 (\eta +6 \xi )\right)\right]\right\}, \label{Eq9} \nn \\
\ea
\ba
p_t(r)&=&\frac{1}{16 \pi G_{\mbox{\footnotesize \ae}} \xi \left(D_1+D_2 r^2\right)^2 \left(D_3+D_4 r^2\right)^2} \left\{2 D_1^2 \xi  \left[D_3 \left(-D_4+D_5+2 D_6 r^2\right)+D_4 D_6 r^4\right]+2 D_1 D_2 \xi  \left[2 D_3^2 \right.\right. \nn \\
&&\left. +D_3 r^2 \left(-D_4+5 D_5+8 D_6 r^2\right)+D_4 r^4 \left(2 D_5+5 D_6 r^2\right)\right]-D_2^2 r^2 \left[D_3^2 (\eta -2 \xi)+D_3 r^2 \left(D_4 (\eta +2 \xi) \right.\right. \nn \\
&&\left.\left.\left. +D_5 (\eta -6 \xi )+D_6 r^2 (\eta -10 \xi )\right)+D_4 r^4 \left(D_5 (\eta -2 \xi )+D_6 r^2 (\eta -6 \xi )\right)\right]\right\}, \label{Eq10} \nn \\
\ea
where $D_1$, $D_2$, $D_3$, $D_4$, $D_5$ and $D_6$ are arbitrary constants. 

The specific choice of $A(r)$ is motivated by the fact that it automatically reproduces the Newtonian potential for a fluid sphere with constant density. Instead, $B(r)$ differs from the Tolmann IV solution for the presence of additional constants, thus making the metric above more general.

The implementation of the junction conditions~\cite{Israel:1966rt} to the exterior vacuum metric amounts to require the continuity of $A(r)$, $B(r)$ and $A'(r)$ at the surface $r=\bar{R}$. By using Eq.~\eqref{Eq2} this implies that also $p_r(r)$ has to be continuous at the surface, {\it i.e.} $p_r(\bar{R})=0$, which leads to the following relation among the arbitrary constants and the radius $\bar{R}$ of the spherical object:
\ba
D_3 = \frac{2 D_1^2 \xi  \left(D_4-D_5-D_6 \bar{R}^2\right)+4 D_1 D_2 \xi \bar{R}^2 \left[D_4-2 \left(D_5+D_6 \bar{R}^2\right)\right]-D_2^2 \bar{R}^4 \left[-2 D_4 \xi +D_5 (\eta +6 \xi )+D_6 \bar{R}^2 (\eta +6 \xi )\right]}{D_2 \left[4 D_1 \xi +D_2 \bar{R}^2 (\eta +4 \xi )\right]}\,. \nn \\
\ea
It is important to stress that, due to the lack of the general explicit vacuum solution in Ho\v rava gravity with a static \ae ther~\cite{Eling:2006df,Eling:2007xh}, one cannot perform a direct analytical calculation of the junction conditions to the exterior spacetime nor an exact inspection of the behavior at infinity. Nevertheless, to leading order in $1/r$, the vacuum solution agrees with the Schwarzschild solution of GR and it is asymptotically flat~\cite{Eling:2006df,Eling:2007xh}, which makes the exterior spacetime anyhow suitable to be continuously joined to the interior one.

In the following Section we will consider some EoS widely used in literature~\cite{Nilsson:2000zg,Herrera:2013fja,Herrera:2014caa,Isayev:2017rci} in order to describe the anisotropic inner fluid distribution of compact objects. In this way we will determine which one is the most suitable for fitting the profiles of the above thermodynamical quantities.

\section{Equations of State}
\label{sec4}

In order to derive the EoS that relate the density to the radial and tangential pressure, we make use of the solutions in Eqs.~\eqref{Eq8}-\eqref{Eq10}, with $G_{\mbox{\footnotesize \ae}}=G_N \left(1-\eta/2\xi \right)$, where $G_N$ is the Newton's constant, which is needed to recover the Newtonian limit~\cite{Carroll:2004ai}, and with $\eta = 2\left(\xi-1\right)$, so that the the post-Newtonian constraints are evaded~\cite{Blas:2010hb,Blas:2011zd,Bonetti:2015oda}. Moreover, we fix the arbitrary constants as done in Ref.~\cite{Vernieri:2017dvi}, {\it i.e.}, $\xi=1.00001$, $D_1=-30$, $D_2=-80$, $D_4=-50$, $D_5=10$, $D_6=-10$, $\bar{R}=0.5$, and we choose the physical units such that $G_N=1$. 
Notice that such choice of the constants is not fine-tuned. Indeed, the phenomenological implications concerning the thermodynamical quantities that will be derived below are qualitatively identical for a very large range of the parameters $D_i$ and $\bar{R}$, and the specific values fixed here are only taken for illustrative purposes.
Then, we calculate the values of the thermodynamical quantities at several fixed radii, as listed in Table~\ref{Tab1}. 

\begin{table*}[!h]
\centering
\begin{tabular}{|c|c|c|c|c|c|}
\hline
\multicolumn{1}{|c|}{  }&
\multicolumn{3}{|c|}{Thermodynamical quantities evaluated at fixed radius}\\
\hline
\hline
Radius ($r$) & Density ($\rho$) & Radial pressure ($p_r$) & Tangential pressure ($p_t$) \\
\hline
\hline
$0$          & $0.3609$		     	& $0.0919$ 			  & $0.0919$         \\[1mm]
$0.05$       & $0.3569$		     	& $0.0897$ 			  & $0.0890$         \\[1mm]
$0.10$       & $0.3453$		   		& $0.0835$    		& $0.0810$         \\[1mm]
$0.15$       & $0.3272$   		  & $0.0739$			  & $0.0691$		     \\[1mm]
$0.20$       & $0.3043$        	& $0.0623$        & $0.0553$         \\[1mm]
$0.25$       & $0.2784$      	  & $0.0496$ 	      & $0.0413$         \\[1mm]
$0.30$ 	     & $0.2511$      	  & $0.0371$        & $0.0286$         \\[1mm]
$0.35$       & $0.2239$         & $0.0256$        & $0.0180$         \\[1mm]
$0.40$       & $0.1978$     	  & $0.0154$        & $0.0098$         \\[1mm]
$0.45$       & $0.1735$         & $0.0068$        & $0.0039$         \\[1mm]
$0.50$       & $0.1514$     	  & $0$       	    & $0.0001$         \\
\hline
\end{tabular}
\caption{Values of the thermodynamical quantities calculated at several fixed radii.}
\label{Tab1}
\end{table*}

\begin{figure}[!h]
\includegraphics[width=0.5\textwidth]{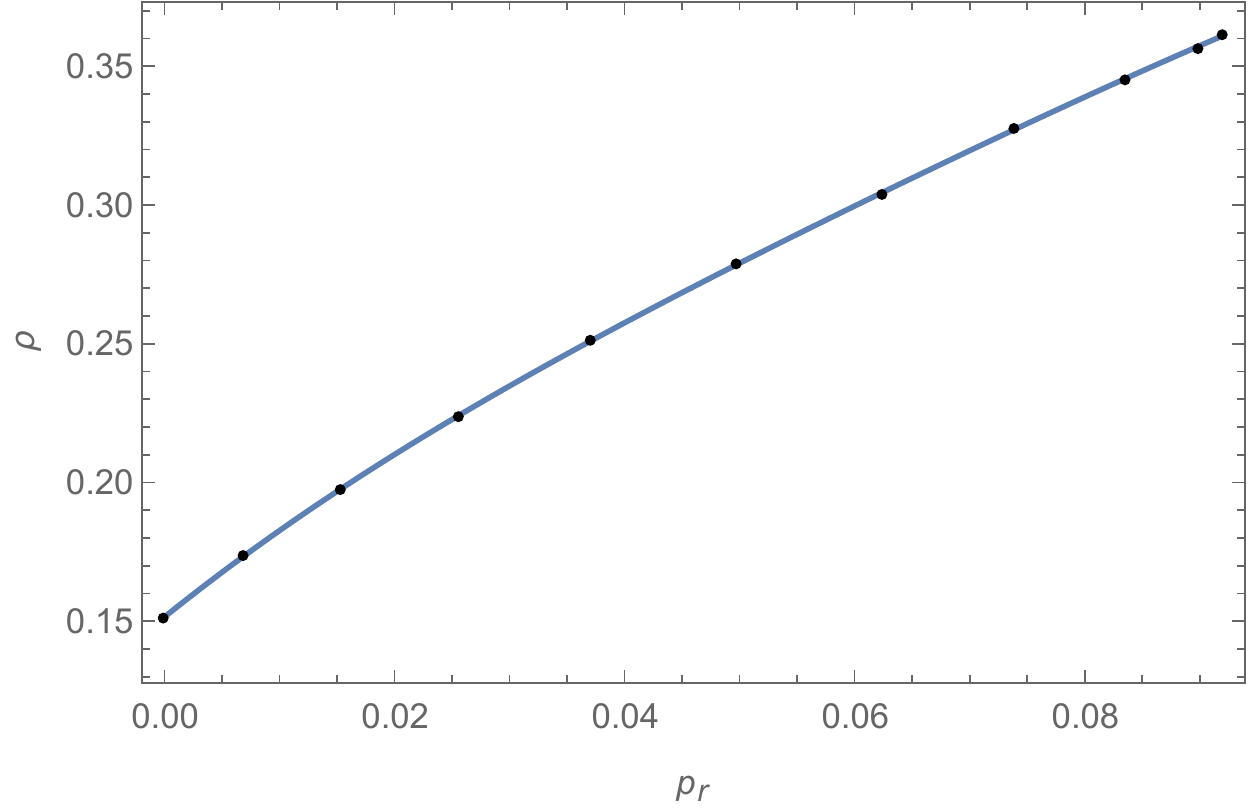}
\caption{The best-fit of the polynomial EoS $\rho=a_0 + a_1 p_r + a_2 p_r^2 + a_3 p_r^3 + a_4 p_r^4$ is shown. The resulting best-fit parameters are: $a_0=0.151515$, $a_1=3.34183$, $a_2=-24.6923$, $a_3=218.84$, and $a_4=-828.709$.}
\label{Figure1}
\end{figure}

\begin{figure}[!h]
\includegraphics[width=0.5\textwidth]{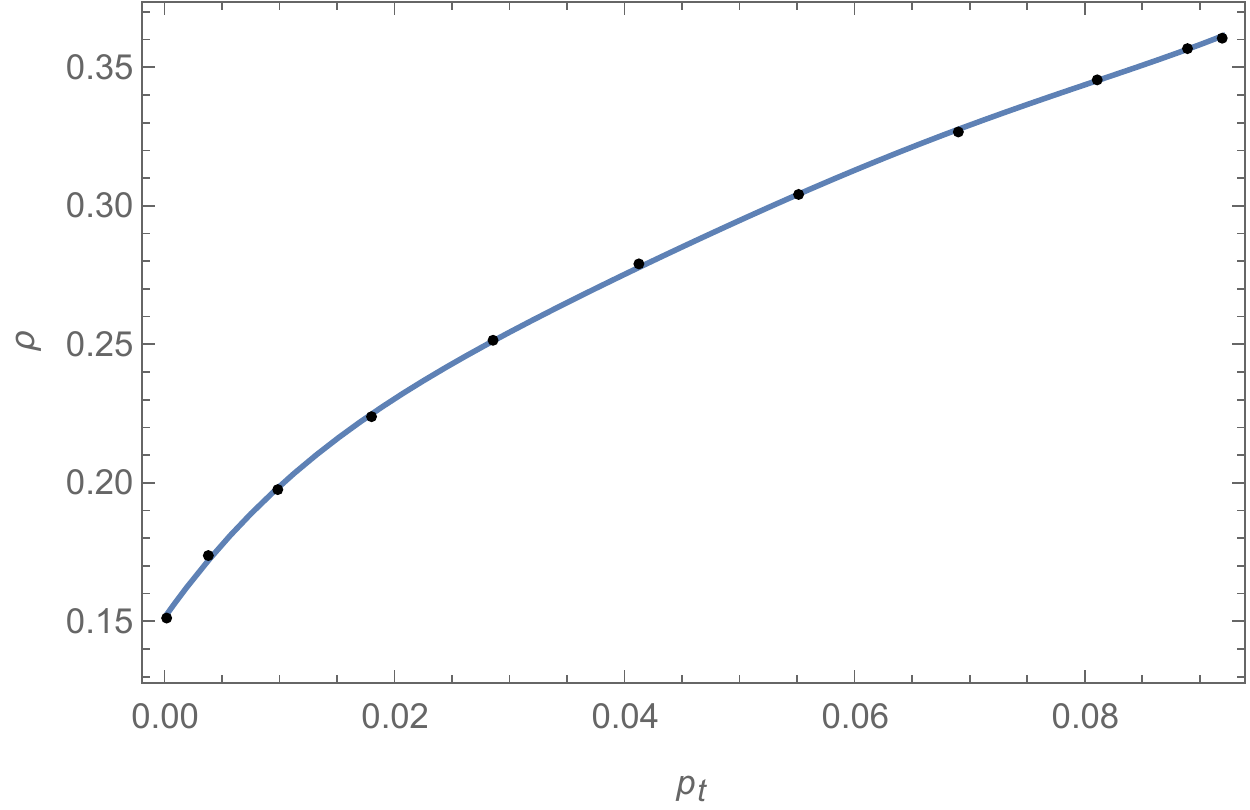}
\caption{The best-fit for the polynomial EoS $\rho=b_0 + b_1 p_t + b_2 p_t^2 + b_3 p_t^3 + b_4 p_t^4 + b_5 p_t^5$ is shown. The best-fit parameters are found to be: $b_0=0.151272$, $b_1=5.92332$, $b_2=-142.096$, $b_3=2637.4$, $b_4=-24862.6$ and $b_5=90318.1$.}
\label{Figure2}
\end{figure}

Let us now consider the scenario in which the EoS $\rho=\rho(p_r)$ and $\rho=\rho(p_t)$ are given by analytic polynomial expressions. So, we can easily calculate the best-fit parameters of the polynomial EoS both for the radial and tangential pressure. The corresponding plots are shown in Figs.~\eqref{Figure1} and~\eqref{Figure2}, respectively. 

In detail, we find that in order to properly fit the EoS for the radial pressure, {\it i.e.} $\rho=\rho(p_r)$, a fourth-order polynomial is needed, that is 
\be
\rho=a_0 + a_1 p_r + a_2 p_r^2 + a_3 p_r^3 + a_4 p_r^4\,,
\ee
where $a_i$ are free parameters. The best-fit parameters are then found to be: $a_0=0.151515$, $a_1=3.34183$, $a_2=-24.6923$, $a_3=218.84$ and $a_4=-828.709$. The resulting profile is plotted in Fig.~\eqref{Figure1}.

Instead, in order to find the best-fit parameters of the EoS related to the tangential pressure, {\it i.e.} $\rho=\rho(p_t)$, we need to use a fifth-order polynomial, which is
\be
\rho=b_0 + b_1 p_t + b_2 p_t^2 + b_3 p_t^3 + b_4 p_t^4 + b_5 p_t^5\,,
\ee
where $b_i$ are free parameters.
In this case, the best-fit parameters are found to be: $b_0=0.151272$, $b_1=5.92332$, $b_2=-142.096$, $b_3=2637.4$, $b_4=-24862.6$, and $b_5=90318.1$ that are used to plot the corresponding EoS in Fig.~\eqref{Figure2}.

\begin{figure}[!h]
\includegraphics[width=0.5\textwidth]{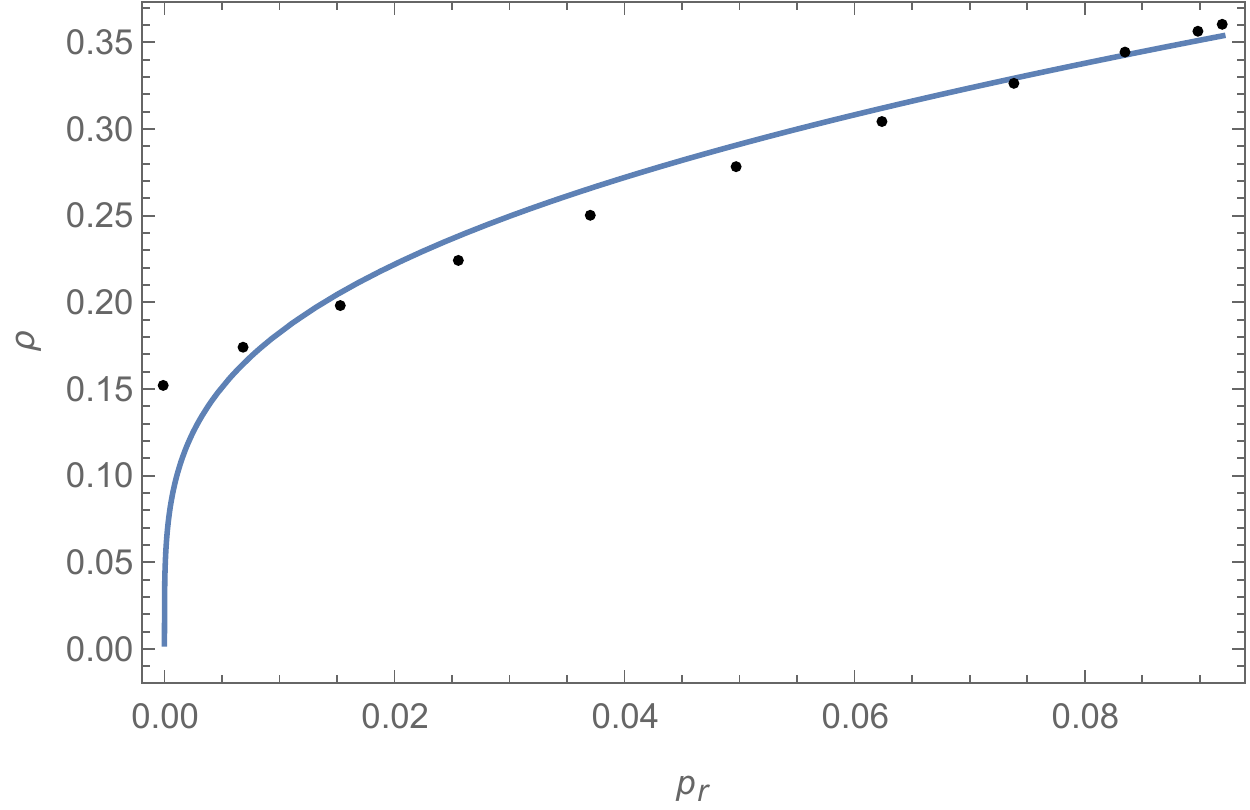}
\caption{The best-fit for the relativistic polytropic EoS $\rho=(p_r/K_r)^{1/\Gamma_r}+p_r/(\Gamma_r-1)$ is shown. The resulting best-fit parameters are: $K_r=6.91$ and $\Gamma_r=3.80$.}
\label{Figure3}
\end{figure}

Let us now consider an alternative possibility widely used in the literature when dealing with EoS, which is to take the relativistic polytropic EoS~\cite{Nilsson:2000zg,Herrera:2013fja,Herrera:2014caa,Isayev:2017rci} in order to reproduce the profile of the density as a function of the radial pressure, 
\be
\rho=(p_r/K_r)^{1/\Gamma_r}+p_r/(\Gamma_r-1),
\label{relpol1}
\ee
where $K_r$ and $\Gamma_r$ are free parameters. 
The plot is shown in Fig.~\eqref{Figure3}, and the corresponding best-fit parameters are found to be: $K_r=6.91$ and $\Gamma_r=3.80$. 

\begin{figure}[!h]
\includegraphics[width=0.5\textwidth]{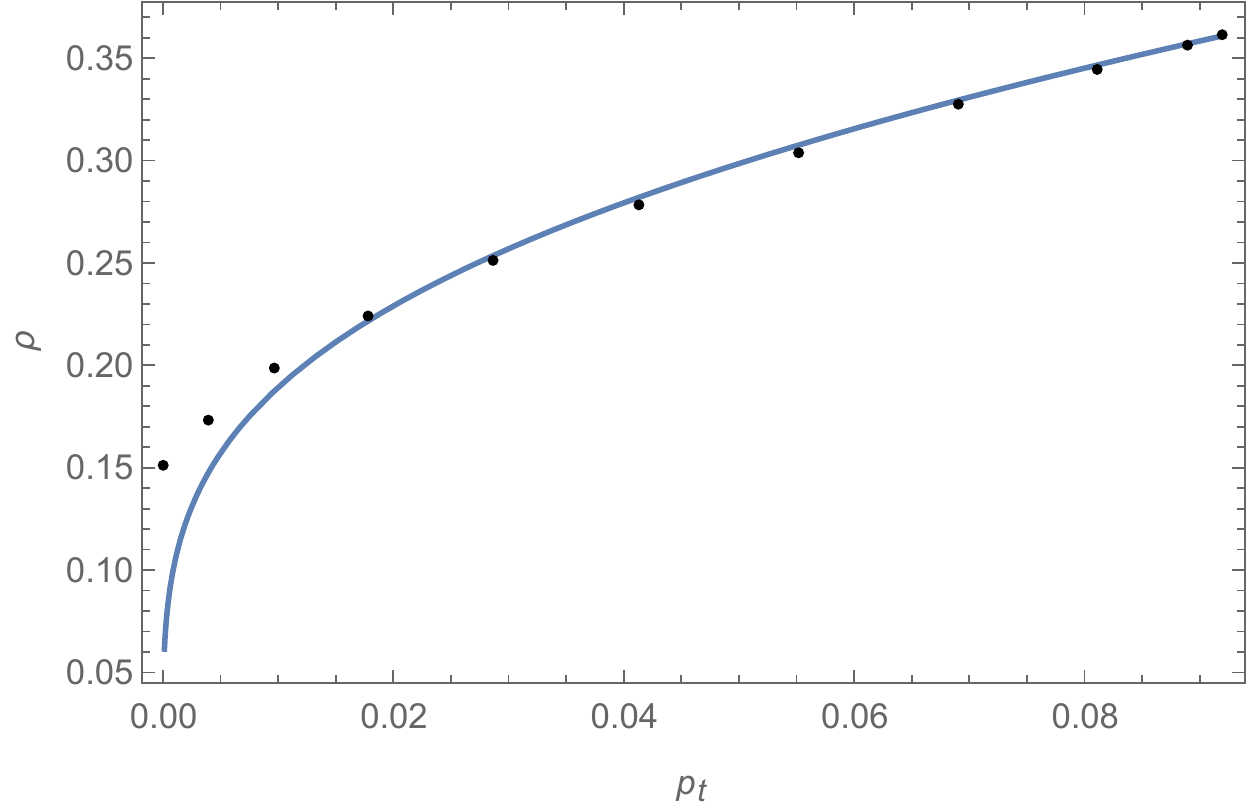}
\caption{The best-fit for the relativistic polytropic EoS $\rho=(p_t/K_t)^{1/\Gamma_t}+p_t/(\Gamma_t-1)$ is shown. The best-fit parameters are found to be: $K_t=6.87$ and $\Gamma_t=3.88$.}
\label{Figure4}
\end{figure}

The same procedure is also applied to $p_t$, looking for an EoS like
\be
\rho=(p_t/K_t)^{1/\Gamma_t}+p_t/(\Gamma_t-1),
\label{relpol2}
\ee
where $K_t$ and $\Gamma_t$ are free parameters. The corresponding plot is shown in Fig.~\eqref{Figure4}, where the best-fit parameters are found to be: $K_t=6.87$ and $\Gamma_t=3.88$. 

As is clear by looking at Figs.~\eqref{Figure3} and~\eqref{Figure4}, the profile of the relativistic polytropic EoS is not satisfactory enough in order to fit the behavior of the density as a function of the radial and tangential pressure. Then, we do a step forward, and by direct inspection, we will try to modify the EoS in Eqs.~\eqref{relpol1}-\eqref{relpol2} through the addition of an extra factor. The EoS that we are going to use will then be called from now on ``modified relativistic polytropic EoS''.

\begin{figure}[!h]
\includegraphics[width=0.5\textwidth]{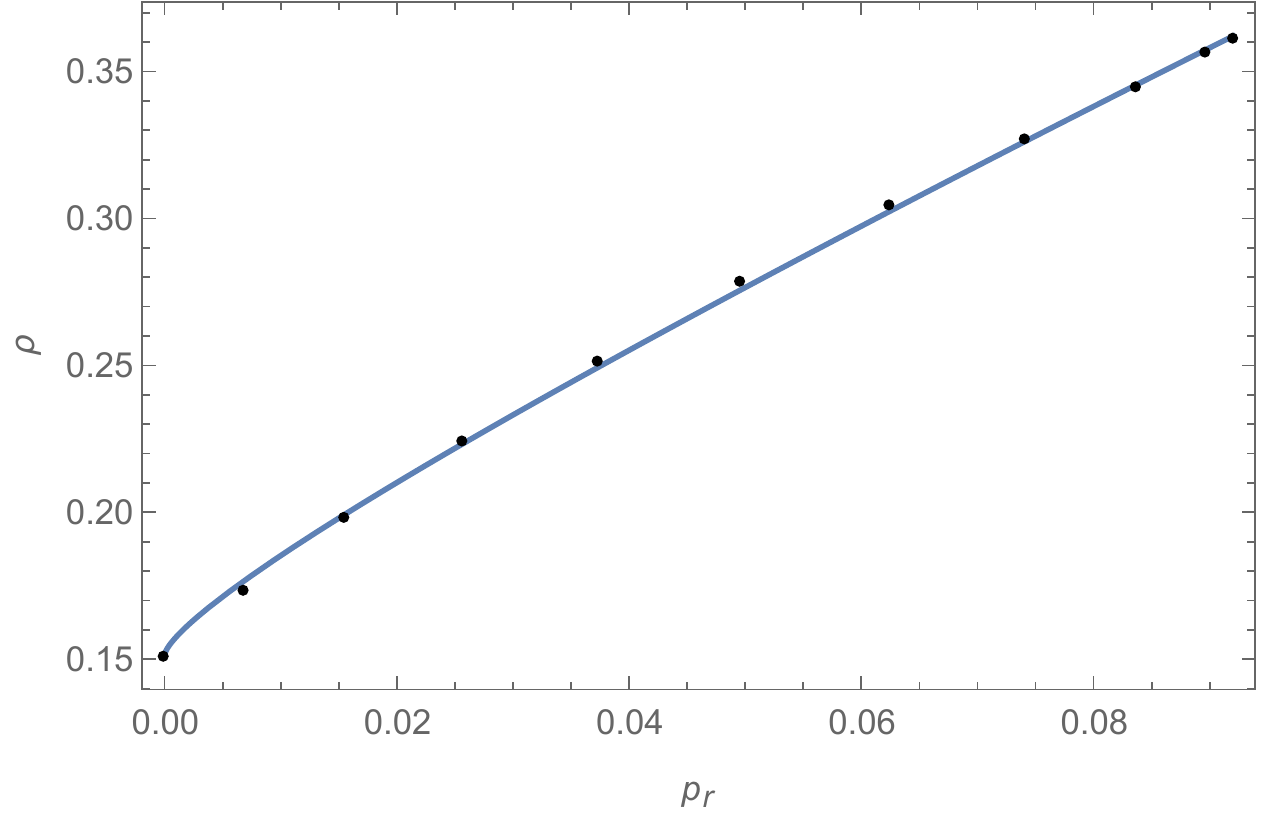}
\caption{The best-fit for the modified relativistic polytropic EoS $\rho=\rho_{0r}+(p_r/K_r')^{1/\Gamma_r'}+p_r/(\Gamma_r'-1)$ is shown. The corresponding best-fit parameters are: $\rho_{0r}=0.1514$, $K_r'=7.38$ and $\Gamma_r'=1.66$.}
\label{Figure5}
\end{figure}

For the radial pressure, we get the best-fit parameters by using the EoS
\be
\rho=\rho_{0r}+(p_r/K_r')^{1/\Gamma_r'}+p_r/(\Gamma_r'-1),
\ee
where $\rho_{0r}$, $K_r'$, and $\Gamma_r'$ are free parameters. The corresponding plot is shown in Fig.~\eqref{Figure5}, and the resulting best-fit parameters are found to be: $\rho_{0r}=0.1514$, $K_r'=7.38$ and $\Gamma_r'=1.66$. 

Notice that in the modified relativistic polytropic EoS for $p_r$ the constant $\rho_{0r}$ has been fixed by hand, in order to account for its exact value as derived from the solution in Eq.~\eqref{Eq8} and reported in Table~\ref{Tab1}.

\begin{figure}[!h]
\includegraphics[width=0.5\textwidth]{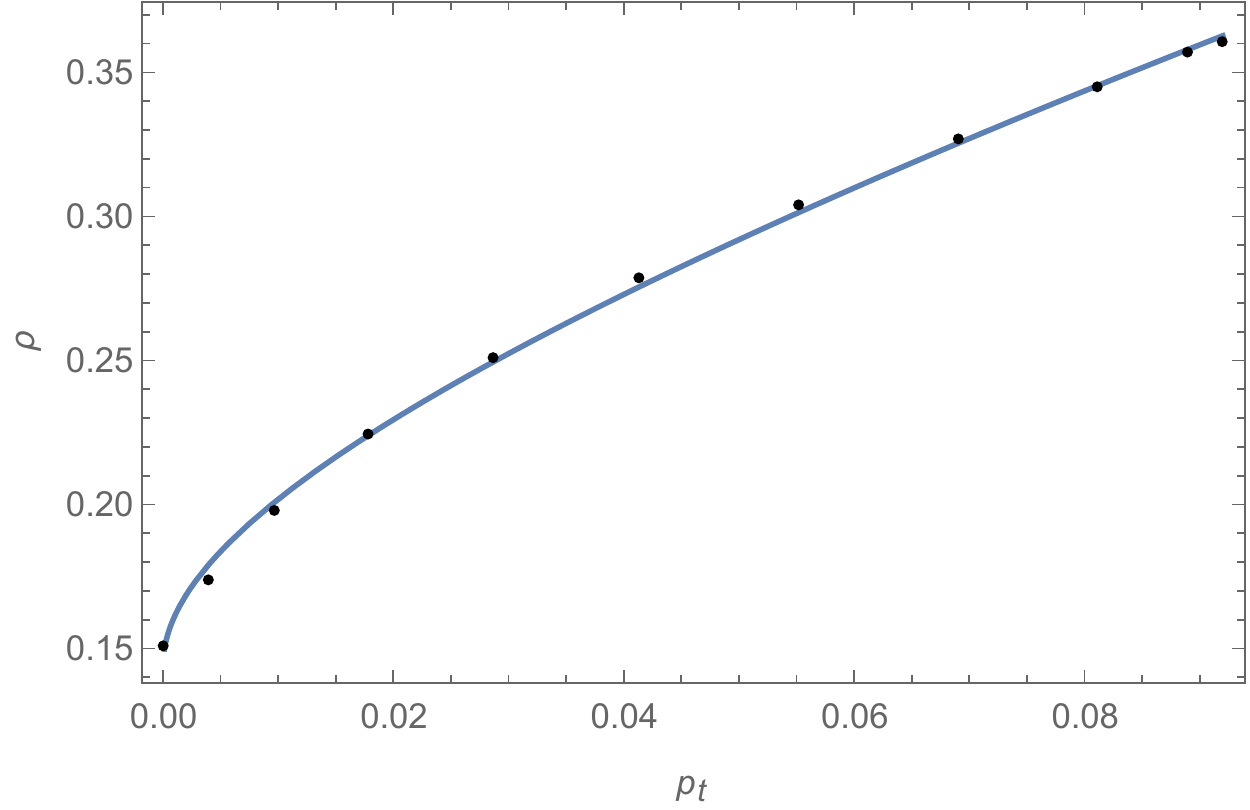}
\caption{The best-fit for the modified relativistic polytropic EoS $\rho=\rho_{0t}+(p_t/K_t')^{1/\Gamma_t'}+p_t/(\Gamma_t'-1)$ is shown. The best-fit parameters are: $\rho_{0t}=0.1428$, $K_t'=6.37$, and $\Gamma_t'=2.16$.}
\label{Figure6}
\end{figure}

For the tangential pressure, the EoS is
\be
\rho=\rho_{0t}+(p_t/K_t')^{1/\Gamma_t'}+p_t/(\Gamma_t'-1),
\ee
where $\rho_{0t}$, $K_t'$, and $\Gamma_t'$ are free parameters. By using the EoS above, one gets the following best-fit parameters: $\rho_{0t}=0.1428$, $K_t'=6.37$ and $\Gamma_t'=2.16$. The profile of the resulting EoS is plotted in Fig.~\eqref{Figure6}.

By looking at Figs.~\eqref{Figure5} and~\eqref{Figure6}, we can observe that the modified relativistic polytropic EoS is able to provide a very good fit for the density as a function of both the radial and tangential pressure.  
Notice that similar results would have been obtained also by using a polytropic EoS of the form $\rho=(p/K)^{1/\Gamma}$~\cite{Nilsson:2000zg,Herrera:2013fja}, both for the radial and tangential pressure. Indeed, it is easy to prove that a correction to the EoS like the one used in the previous case would be again necessary in order to properly describe the relation existing among the various thermodynamical quantities.
 
The analysis presented above clearly indicates that, in particular in the context of a theory like Ho\v rava gravity, which is a candidate to be UV complete at quantum gravity scales, the standard microscopic thermodynamical description of the inner fluid components needs to be consistently revised in order to account for the new phenomenological implications predicted by the theory.

\section{Conclusions}
\label{sec5}

We have considered the viable analytic and exact interior solutions by assuming spherical symmetry and with the addition of anisotropic fluids in the context of Ho\v rava gravity and Einstein-\ae ther theory~\cite{Vernieri:2017dvi}, and we have derived the corresponding EoS. We started by taking the explicit solutions for the density and the radial and tangential pressure, and we first proceeded by obtaining their corresponding values at some fixed radii. Then, by means of those, we have performed a polynomial fit for the EoS, $\rho=\rho(p_r)$ and $\rho=\rho(p_t)$, getting the exact values of the parameters characterizing the polynomial expressions of the two EoS, respectively.
After that, we also used the functional form of the relativistic polytropic EoS widely used in the literature~\cite{Nilsson:2000zg,Herrera:2013fja,Herrera:2014caa,Isayev:2017rci}, and we showed that it is not accurate enough to describe the relation between density and the radial and tangential pressure. By means of a direct inspection, we then noticed that a correction to the relativistic polytropic EoS is needed in order to get a satisfactory fit for the density as a function of the radial and tangential pressure. In this case, the modification has consisted of the addition of a constant factor to the standard relativistic polytropic EoS. We referred to this new class of EoS as modified relativistic polytropic EoS. Moreover, we have also commented on the fact that the same conclusion qualitatively applies to the common polytropic EoS widely used in the literature~\cite{Nilsson:2000zg,Herrera:2013fja}, too.
To conclude, this result suggests that the standard relativistic EoS is definitely not appropriate to describe the inner spacetime of compact objects in the low-energy limit of Ho\v rava gravity with a static \ae ther. This makes it even more challenging to study the predictions of the theory about the microscopic thermodynamical description of the inner spacetime, since the standard picture has been proven to be definitely not accurate enough in the case under study.

\vspace{0.3cm}
{\bf Acknowledgments:} The author would like to thank Antonio De Felice for enlightening discussions that led to the initiation of this project.
The author was supported by the Funda\c{c}\~{a}o para a Ci\^{e}ncia e Tecnologia through Project No. IF/00250/2013 and acknowledge financial support provided under the European Union's H2020 ERC Consolidator Grant ``Matter and strong-field gravity: New frontiers in Einstein's theory'' grant agreement No. MaGRaTh646597.



\begin{thebibliography}{99}

\bibitem{Horava:2009uw} 
  P.~Horava,
  Phys.\ Rev.\ D {\bf 79}, 084008 (2009)
  [arXiv:0901.3775 [hep-th]].

\bibitem{Blas:2009qj} 
  D.~Blas, O.~Pujolas and S.~Sibiryakov,
  Phys.\ Rev.\ Lett.\  {\bf 104}, 181302 (2010)
  [arXiv:0909.3525 [hep-th]].

\bibitem{Barvinsky:2015kil} 
  A.~O.~Barvinsky, D.~Blas, M.~Herrero-Valea, S.~M.~Sibiryakov and C.~F.~Steinwachs,
  Phys.\ Rev.\ D {\bf 93}, no. 6, 064022 (2016)
  [arXiv:1512.02250 [hep-th]].
	
\bibitem{Bellorin:2016wsl} 
  J.~Bellor\'in and A.~Restuccia,
  Phys.\ Rev.\ D {\bf 94}, no. 6, 064041 (2016)
  [arXiv:1606.02606 [hep-th]].
	
\bibitem{Barvinsky:2017zlx} 
  A.~O.~Barvinsky, D.~Blas, M.~Herrero-Valea, S.~M.~Sibiryakov and C.~F.~Steinwachs,
  JHEP {\bf 1807}, 035 (2018)
  [arXiv:1705.03480 [hep-th]].

\bibitem{Sotiriou:2009gy} 
  T.~P.~Sotiriou, M.~Visser and S.~Weinfurtner,
  Phys.\ Rev.\ Lett.\  {\bf 102}, 251601 (2009)
  [arXiv:0904.4464 [hep-th]].

\bibitem{Weinfurtner:2010hz} 
  S.~Weinfurtner, T.~P.~Sotiriou and M.~Visser,
  J.\ Phys.\ Conf.\ Ser.\  {\bf 222}, 012054 (2010)
  [arXiv:1002.0308 [gr-qc]].

\bibitem{Vernieri:2011aa} 
  D.~Vernieri and T.~P.~Sotiriou,
  Phys.\ Rev.\ D {\bf 85}, 064003 (2012)
  [arXiv:1112.3385 [hep-th]].

\bibitem{Vernieri:2012ms} 
  D.~Vernieri and T.~P.~Sotiriou,
  J.\ Phys.\ Conf.\ Ser.\  {\bf 453}, 012022 (2013)
  [arXiv:1212.4402 [hep-th]].

\bibitem{Vernieri:2015uma} 
  D.~Vernieri,
  Phys.\ Rev.\ D {\bf 91}, no. 12, 124029 (2015)
  [arXiv:1502.06607 [hep-th]].

\bibitem{Berglund:2012bu} 
  P.~Berglund, J.~Bhattacharyya and D.~Mattingly,
  Phys.\ Rev.\ D {\bf 85}, 124019 (2012)
  [arXiv:1202.4497 [hep-th]].

\bibitem{Sotiriou:2014gna} 
  T.~P.~Sotiriou, I.~Vega and D.~Vernieri,
  Phys.\ Rev.\ D {\bf 90}, no. 4, 044046 (2014)
  [arXiv:1405.3715 [gr-qc]].

\bibitem{Lu:2009em} 
  H.~Lu, J.~Mei and C.~N.~Pope,
  Phys.\ Rev.\ Lett.\  {\bf 103}, 091301 (2009)
  [arXiv:0904.1595 [hep-th]].

\bibitem{Cai:2009pe} 
  R.~G.~Cai, L.~M.~Cao and N.~Ohta,
  Phys.\ Rev.\ D {\bf 80}, 024003 (2009)
  [arXiv:0904.3670 [hep-th]].

\bibitem{Park:2009zra} 
  M.~i.~Park,
  JHEP {\bf 0909}, 123 (2009)
  [arXiv:0905.4480 [hep-th]].

\bibitem{Eling:2006df} 
  C.~Eling and T.~Jacobson,
  Class.\ Quant.\ Grav.\  {\bf 23}, 5625 (2006)
  Erratum: [Class.\ Quant.\ Grav.\  {\bf 27}, 049801 (2010)]
  [gr-qc/0603058].
	
\bibitem{Eling:2007xh} 
  C.~Eling, T.~Jacobson and M.~Coleman Miller,
  Phys.\ Rev.\ D {\bf 76}, 042003 (2007)
  Erratum: [Phys.\ Rev.\ D {\bf 80}, 129906 (2009)]
  [arXiv:0705.1565 [gr-qc]].

\bibitem{Vernieri:2017dvi} 
  D.~Vernieri and S.~Carloni,
  EPL {\bf 121}, no. 3, 30002 (2018)
  [arXiv:1706.06608 [gr-qc]].

\bibitem{Herrera:1997plx} 
  L.~Herrera and N.~O.~Santos,
  Phys.\ Rept.\  {\bf 286}, 53 (1997).

\bibitem{Harko:2002db} 
  T.~Harko and M.~K.~Mak,
  Annalen Phys.\  {\bf 11}, 3 (2002)
  [gr-qc/0302104].

\bibitem{Ozel:2016oaf} 
  F.~Ozel and P.~Freire,
  Ann.\ Rev.\ Astron.\ Astrophys.\  {\bf 54}, 401 (2016)
  [arXiv:1603.02698 [astro-ph.HE]];

\bibitem{Jacobson:2000xp} 
  T.~Jacobson and D.~Mattingly,
  Phys.\ Rev.\ D {\bf 64}, 024028 (2001)
  [gr-qc/0007031].

\bibitem{Nilsson:2000zg} 
  U.~S.~Nilsson and C.~Uggla,
  Annals Phys.\  {\bf 286}, 292 (2001)
  [gr-qc/0002022].

\bibitem{Herrera:2013fja} 
  L.~Herrera and W.~Barreto,
  Phys.\ Rev.\ D {\bf 88}, no. 8, 084022 (2013)
  [arXiv:1310.1114 [gr-qc]].

\bibitem{Herrera:2014caa} 
  L.~Herrera, A.~Di Prisco, W.~Barreto and J.~Ospino,
  Gen.\ Rel.\ Grav.\  {\bf 46}, no. 12, 1827 (2014)
  [arXiv:1410.6636 [gr-qc]].
	
\bibitem{Isayev:2017rci} 
  A.~A.~Isayev,
  Phys.\ Rev.\ D {\bf 96}, no. 8, 083007 (2017)
  [arXiv:1801.03745 [gr-qc]].

\bibitem{Jacobson:2010mx} 
  T.~Jacobson,
  Phys.\ Rev.\ D {\bf 81}, 101502 (2010)
  Erratum: [Phys.\ Rev.\ D {\bf 82}, 129901 (2010)]
  [arXiv:1001.4823 [hep-th]].

\bibitem{Blas:2010hb} 
  D.~Blas, O.~Pujolas and S.~Sibiryakov,
  JHEP {\bf 1104}, 018 (2011)
  [arXiv:1007.3503 [hep-th]].

\bibitem{Tolman:1939jz} 
  R.~C.~Tolman,
  Phys.\ Rev.\  {\bf 55}, 364 (1939).

\bibitem{Israel:1966rt} 
  W.~Israel,
  Nuovo Cim.\ B {\bf 44S10}, 1 (1966)
  [Nuovo Cim.\ B {\bf 44}, 1 (1966)]
  Erratum: [Nuovo Cim.\ B {\bf 48}, 463 (1967)].

\bibitem{Carroll:2004ai} 
  S.~M.~Carroll and E.~A.~Lim,
  Phys.\ Rev.\ D {\bf 70}, 123525 (2004)
  [hep-th/0407149].
	
\bibitem{Blas:2011zd} 
  D.~Blas and H.~Sanctuary,
  Phys.\ Rev.\ D {\bf 84}, 064004 (2011)
  [arXiv:1105.5149 [gr-qc]].

\bibitem{Bonetti:2015oda} 
  M.~Bonetti and E.~Barausse,
  Phys.\ Rev.\ D {\bf 91}, 084053 (2015)
  Erratum: [Phys.\ Rev.\ D {\bf 93}, 029901 (2016)]
  [arXiv:1502.05554 [gr-qc]].
																		
\end{thebibliography}
\end{document}